\documentclass[12pt,onecolumn,draftcls]{IEEEtran}
\usepackage{bbm}
\usepackage{dsfont}
\usepackage{yfonts}
\usepackage{bm,cleveref}
\usepackage{amsmath}
\usepackage{amssymb}
\usepackage{amsfonts}
\usepackage{mathrsfs} 
\usepackage{stfloats}
\usepackage{amsthm}
\usepackage{subfigure}
\usepackage{graphicx}
\usepackage{epstopdf,amssymb,amsmath}
\usepackage{multicol}
\usepackage[noadjust]{cite}
\usepackage{setspace}
\usepackage{stfloats,subfigure}
\usepackage{midfloat}
\usepackage[normal]{threeparttable}
\usepackage{amsthm}
\usepackage{cuted}
\usepackage{cases,subeqnarray}
\usepackage{bm,multirow,bigstrut}
\usepackage{textcomp}
\usepackage{latexsym,bm}
\usepackage{booktabs,changebar}
\usepackage{xcolor}
\usepackage{mathtools}
\usepackage{dsfont}
\usepackage{extarrows}

\newtheorem{thm}{Theorem}
\newtheorem{coro}{Corollary}

\ifCLASSOPTIONcompsoc
\usepackage[nocompress]{cite}
\else
\usepackage{cite}
\fi

\IEEEoverridecommandlockouts
\begin{document}
\title{Spectral and Energy Efficiency of Cell-Free Massive MIMO Systems with Hardware Impairments}
\author{Jiayi~Zhang,~
        Yinghua Wei,~
        Emil~Bj\"{o}rnson,~
        Yu Han,
        and Xu Li

\thanks{This work was supported in part by the National Natural Science Foundation of China (Grant No. 61601020) and the Fundamental Research Funds for the Central Universities (Grant Nos. 2016RC013, 2017JBM319, and 2016JBZ003). The work of E.   Bj\"{o}rnson was supported by ELLIIT and SSF, Y. Han was supported in part by the National Science Foundation (NSFC) for Distinguished Young Scholars of China with Grant 61625106.}
\thanks{J. Zhang, Y. Wei, and X. Li are with the School of Electronic and Information Engineering, Beijing Jiaotong University, Beijing 100044, P. R. China (e-mail: jiayizhang@bjtu.edu.cn).}
\thanks{E. Bj\"{o}rnson is with the Department of Electrical Engineering (ISY), Link\"{o}ping University, Link\"{o}ping, Sweden.}
\thanks{Y. Han is with the National Mobile Communications Research Laboratory, Southeast University, Nanjing 210096, P. R. China.}}

\maketitle

\begin{abstract}
Cell-free massive multiple-input multiple-output (MIMO), with a large number of distributed access points (APs) that jointly serve the user equipments (UEs), is a promising network architecture for future wireless communications. To reduce the cost and power consumption of such systems, it is important to utilize low-quality transceiver hardware at the APs. However, the impact of hardware impairments on cell-free massive MIMO has thus far not been studied. In this paper, we take a first look at this important topic by utilizing well-established models of hardware distortion and deriving new closed-form expressions for the spectral and energy efficiency. These expressions provide important insights into the practical impact of hardware impairments and also how to efficiently deploy cell-free systems. Furthermore, a novel hardware-quality scaling law is presented. It proves that the impact of hardware impairments at the APs vanish as the number of APs grows. Numerical results validate that cell-free massive MIMO systems are inherently resilient to hardware impairments.
\end{abstract}

\section{Introduction}
Massive MIMO is a cellular technology that equips each cell with a large number of antennas to spatially multiplex many UEs on the same time-frequency resource. It is a key technology for the fifth generation (5G) cellular networks \cite{Liu2017Energy,Hoydis2013a,Boccardi2014Five}, since it can offer high spectral and energy efficiency under practical conditions, which is necessary to keep up with the tremendous growing demand for wireless communications \cite{Bjornson2016Massive}. There are two topologies for cellular massive MIMO deployment: a co-located antenna array in the cell center or multiple geographically distributed antenna arrays \cite{bjornson2015massive}. Both topologies rely on having cells that each serves an exclusive set of UEs.

Instead of relying on cells, a network MIMO approach can be taken where geographically distributed APs are jointly serving all the UEs \cite{Venkatesan2007Network}. Cell-free massive MIMO is the latest form of network MIMO, where a massive number of single-antenna APs are deployed to phase-coherently and simultaneously serve a much smaller number of UEs, distributed over a wide area. To make the network operation scalable, a time-division duplex (TDD) protocol is used, where each AP only utilizes locally estimated channels and only data signals are distributed over the backhaul \cite{Bjornson2010Cooperative}. What makes cell-free massive MIMO different from classic network MIMO is the analytical approach, borrowed from cellular massive MIMO, which enables ergodic capacity analysis and efficient power control.
In \cite{Nayebi2017Precoding}, the authors developed a max-min power control algorithm combined with linear zero-forcing precoders for cell-free massive MIMO. In \cite{Ngo2015Cell,Ngo2016Cell}, it was indicated that cell-free massive MIMO can give a 5--10 fold gain in throughput over uncoordinated small cell systems, taking the effects of imperfect channel state information (CSI), pilot assignment and power control into consideration.
In \cite{nayebi2016performance}, the authors presented an asymptotic approximation of the signal-to-interference-plus-noise ratio (SINR) of the minimum mean-square error (MMSE) receiver. In \cite{nguyen2017energy,ngo2017total}, the authors developed a novel low-complexity power control technique with zero-forcing precoding to maximize the energy efficiency (EE) of cell-free massive MIMO.

The aforementioned works on cell-free massive MIMO assume that the transceiver hardware of the APs is perfect. Considering that the energy consumption and deployment cost increase rapidly with the number of APs, cell-free massive MIMO systems preferably use low-cost components, which are prone to hardware impairments. It has been proved that the detrimental impact of hardware impairments at the APs of co-located massive MIMO systems vanishes asymptotically as the number of antennas grows large \cite{bjornson2015massive,zhang2016spectral,Zhang2017Performance,Zhang2016On,Zhang2016Achievable}, while the impairments at the UEs remain \cite{Bjornson2014Massive,Zhang2016Achievable}. Whether these properties carry over to cell-free massive MIMO systems is a practically important open question that we will answer.

In this paper, we quantitatively investigate the uplink performance of a cell-free massive MIMO with hardware impairments at APs and UEs. The main contributions are:
\begin{itemize}
\item We derive a closed-form expression for the uplink spectral efficiency (SE) of cell-free massive MIMO systems with transceiver hardware impairments. This expression explicitly reveals how hardware impairments at the UEs and APs affect the SE.
\item We obtain a useful hardware-quality scaling law, which establishes a precise relationship between the number of APs and the hardware quality factors. We prove that the impact of the AP hardware quality vanishes as the number of APs grows large.
\item A closed-form EE expression is derived to show the optimal number of APs for different hardware qualities.
\end{itemize}


\section{System Model}\label{se:model}

\begin{figure}[htbp]
\centering
\includegraphics[scale=0.6]{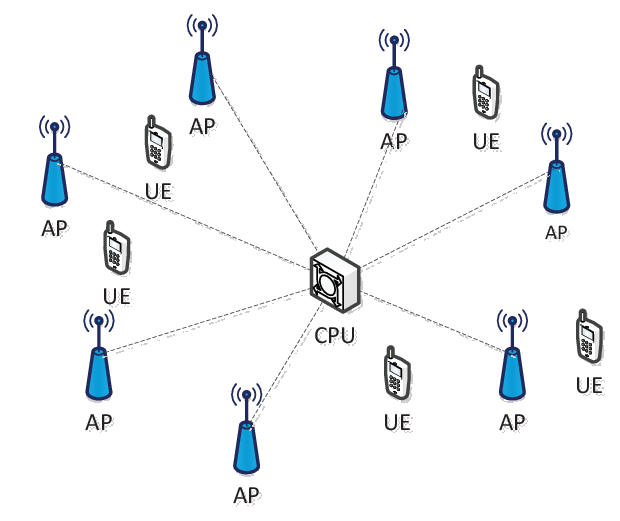} \vspace{-2mm}
\caption{Illustration of a cell-free massive MIMO system.}
\label{model} \vspace{-3mm}
\end{figure}

We consider a cell-free massive MIMO system with $M$ APs and $K$ UEs that are served on the same time-frequency resource. All APs and UEs are equipped with a single antenna in this paper, and they are distributed over a wide area. The APs connect to a central processing unit (CPU) via backhaul links; see Fig~\ref{model} for an illustration.
We consider the classic block fading model \cite{Marzetta2016a,Ngo2016Cell}, where each coherence interval consists of three phases: uplink training, uplink data transmission, and downlink data transmission. In the uplink training, the UEs send pilot sequences and each AP estimates its channel to each UE. The obtained channel estimates are later used to detect the signals transmitted from the UEs in the uplink. In this paper, we only consider the uplink.

The channel coefficient $g_{mk}$ between AP $m$ and UE $k$ is
 \begin{equation}\label{gmk}
 {g_{mk}} \sim \mathcal{CN}(0,\beta _{mk})
 \end{equation}
 for $m = 1,\dots,M$, $k = 1,\dots,K$. The variance ${\beta _{mk}} = \mathbb{E}\{ {{{\left| {{g_{mk}}} \right|}^2}}\}$ denotes the large-scale fading (including path loss and shadowing) and the random distribution models the Rayleigh small-scale fading.

In the uplink data transmission, the $k$th UE first multiplies its information symbol $q_{k}\sim\mathcal{CN}(0,1)$  by a power control coefficient ${\sqrt {{\gamma_k}} ,}$ ($0 \leqslant {\gamma_k} \leqslant 1$). Then, all UEs simultaneously transmit their data to the APs.
In prior works, the received signal at the $m$th AP has been given as
\begin{equation} \label{yum_original}
{y_{um}} = \sum_{k = 1}^K {g_{mk}} \sqrt {\rho _u{\gamma_k}} {q_k} + w_{um},
\end{equation}
where  ${{\rho _ {u}}}$ denotes the maximum transmit power of a UE and ${w _{um}} \sim \mathcal{CN}\left( {0,{\sigma ^2}} \right)$ is the additive white Gaussian noise (AWGN). This model implicitly assumes perfect transceiver hardware.
In practice, the transceiver hardware of the APs and UEs suffer from hardware impairments, which distort the transmitted and received signals. To analyze the joint impact of all kinds of hardware distortion on the communication performance, we use the well-established model from \cite{Bjornson2014Massive}, which is based on measurements \cite{studer2010mimo}. The main characteristic of this model is that the signal power is reduced by a factor $\kappa$ and then additive noise is added with a power that corresponds to the removed signal power.
Applying this model to \eqref{yum_original}, the received signal at the $m$th AP is instead given by
\begin{eqnarray} \label{yum}
{y_{um}} = \sum\limits_{k = 1}^K {{\sqrt{{\kappa _r}}}{g_{mk}}\left( {\sqrt {\rho _u{\gamma_k}{{\kappa _t}}} {q_k} + {\eta _{kt}}} \right) + {\eta _{mr}} + {w_{um}}},
\end{eqnarray}
where $ \kappa _ {t} $ and $ {\kappa _ {r}} $ are the hardware quality factors of the transmitter and receiver, respectively. These are parameters between 0 and 1, where $ \kappa _ {t} = {\kappa _ {r}} =1$ is perfect hardware and $ \kappa _ {t} = {\kappa _ {r}} =0$ is useless hardware that turns everything into distortion.
%
%
Measurements (e.g., \cite{studer2010mimo}) have suggested that 
\begin{eqnarray} \label{HIt}
{\eta _{kt}} \sim \mathcal{CN}\left( {0,\left(1-{\kappa _t}\right)\rho _u{\gamma_k}} \right),
\end{eqnarray}
\begin{eqnarray} \label{HIr}
 {\eta _{mr}} \left| \left\{ {{g_{mk}}} \right\} \right.\sim \mathcal{CN}\left( {0, { \left(1-{\kappa _r}\right) }\rho _u \sum\limits_{k = 1}^K {{\gamma_k}{{\left| {{g_{mk}}} \right|}^2}} } \right),
\end{eqnarray}
where \eqref{HIr} is the conditional distribution given the set of channel realizations $\{ {{g_{mk}}} \}$ in a coherence interval.

The channel estimation at the APs is based on uplink pilots from the UEs. In the uplink training phase, let ${\tau }$ and $\rho _p$ denote the pilot length and the transmit power of each pilot symbol, respectively. The $k$th UE sends its pilot sequences $\sqrt {{\tau }} {\bm{\varphi} _k} \in {\mathbb{C}^{{\tau }\times 1}}$, which satisfies ${\left\| {{\bm{\varphi} _k}} \right\|^2}{\text{  =  1}}$.
Based on the system model (\ref{yum}), the received pilot vector ${{\mathbf{y}}_{pm}} \in \mathbb{C}^{\tau \times 1}$ at the $m$th AP is
\begin{eqnarray} \label{ypm}
{{\mathbf{y}}_{pm}} = \sum\limits_{k = 1}^K {{\sqrt{{\kappa _r}}g_{mk}}\left( {\sqrt {{\tau }\rho _p{\kappa _t}} {{\bm{\varphi }}_k} + {\bm\eta _{kt}}} \right)}  + {\bm{\eta }}_{mr} + {{\mathbf{w}}_{pm}},
\end{eqnarray}
where ${{\mathbf{w}}_{pm}} \sim \mathcal{CN}\left( {\mathbf{0},{{\sigma }}^2 \mathbf{I}_\tau} \right)$ is a vector of AWGN.
Assuming that the distortion is independent between samples in the coherence interval, the transmitter distortion vector is ${{\bm{\eta }}_{kt}} \sim \mathcal{CN}\left( {\mathbf{0},\rho_p \left(1-{ {\kappa _t} }\right){{\rm \mathbf{I}}_{{\tau } }}} \right)$ and the receiver distortion vector is ${{\bm{\eta }}_{mr}}\left| \left\{ {{g_{mk}}} \right\} \right. \sim \mathcal{CN}\left( {\mathbf{0},\rho_p \left(1-{{ {\kappa _r} }}\right) \sum_{k=1}^K{{\left| {{g_{mk}}} \right|}^2}{{\rm \mathbf{I}}_{{\tau  }  }}} \right)$.

Having completed the pilot transmission, the first step towards estimating the channel $g_{mk}$ from UE $k$ is to perform a despreading operation \cite{Marzetta2016a}. More precisely, the AP takes the inner product between $\bm{\varphi} _k$
and ${\mathbf{y}}_{pm}$ to obtain
${{\overset{\lower0.5em\hbox{$\smash{\scriptscriptstyle\smile}$}}{y} }_{p,mk}} = {\bm{\varphi }}_k^H{{\mathbf{y}}_{pm}} .$
The linear MMSE (LMMSE) estimate of $g_{mk}$ based on ${{\overset{\lower0.5em\hbox{$\smash{\scriptscriptstyle\smile}$}}{y} }_{p,mk}}$ is then given by
\begin{eqnarray}\label{hatgmk}
{{\hat g}_{mk}} = \frac{{\mathbb{E}\{ {\mathord{\buildrel{\lower3pt\hbox{$\scriptscriptstyle\smile$}}
\over y} _{p,mk}^*}{g_{mk}}\} }}{{\mathbb{E}\left\{ {{{\left| {{{\overset{\lower0.5em\hbox{$\smash{\scriptscriptstyle\smile}$}}{y} }_{p,mk}}} \right|}^2}} \right\}}}{{\overset{\lower0.5em\hbox{$\smash{\scriptscriptstyle\smile}$}}{y} }_{p,mk}} = {c_{mk}}{{\overset{\lower0.5em\hbox{$\smash{\scriptscriptstyle\smile}$}}{y} }_{p,mk}},
\end{eqnarray}
where $c_{mk}$ is given in 
\begin{align}\label{cmk}
{c_{mk}} \triangleq \frac{{\sqrt {{\tau  }\rho _p {{\kappa _r}{\kappa _t}}} {\beta _{mk}}}}{{\rho _p \sum\limits_{k' = 1}^K {{\beta _{mk'}}\left( {{{\kappa _r}{\kappa _t}} \tau {{\left| {{\bm{\varphi }}_k^H{{\bm{\varphi }}_{k'}}} \right|}^2} + {{{\left(1-\kappa _r \kappa _t\right)}}}} \right)} + \sigma^2}}
\end{align} 
and
\begin{eqnarray}\label{gamamk}
{\lambda _{mk}} \triangleq \mathbb{E}\left\{ {{{\left| {{{\hat g}_{mk}}} \right|}^2}} \right\} = \sqrt {{\tau  }\rho _p {{\kappa _r}{\kappa _t}}} {\beta _{mk}}{c_{mk}}.
\end{eqnarray}

During uplink data transmission, the $m$th AP multiplies its received signal ${{y_{um}}}$ in \eqref{yum} with the conjugate of the LMMSE estimate ${{{\hat g}_{mk}}}$.
Then, each AP sends its obtained quantity ${\hat g_{mk}^*{y_{um}}}$ to the CPU via the backhaul. The combined received signal at the CPU is the maximum ratio combined scalar
\begin{align} \label{ruk}
  {r_{uk}}  = \sum\limits_{m = 1}^M {\hat g_{mk}^*{y_{um}}} .
\end{align}

\section{Performance Analysis}\label{se:uplink}

The received signal in \eqref{ruk} can be expanded as
\begin{align} \label{ruk1}
  {r_{uk}} &= \underbrace {\sqrt {\rho _{u} {\gamma_k}{{\kappa _r}{\kappa _t}}} {q_k}\sum\limits_{m = 1}^M {\hat g_{mk}^*{g_{mk}}} }_{k{\text{th UE's signal}}} + \underbrace {\sum\limits_{m = 1}^M {\hat g_{mk}^*{w_{um}}} }_{{\text{compound noise}}}\hfill \notag\\
  &+ \underbrace {\sqrt {\rho _{u} {{\kappa _r}{\kappa _t}}} \sum\limits_{m = 1}^M {\sum\limits_{k' \ne k}^K {\sqrt {{\gamma_{k'}}} {q_{k'}}\hat g_{mk}^*{g_{mk'}}} } }_{{\text{inter-UE interference}}}  \hfill \notag\\
  &+ \underbrace {\sum\limits_{m = 1}^M {\sum\limits_{k' = 1}^K {\sqrt {{\kappa _r}} }{\hat g_{mk}^*{g_{mk'}}{\eta _{k't}}  + \sum\limits_{m = 1}^M {\hat g_{mk}^*{\eta _{mr}}} } } }_{{\text{hardware impairments}}}.
\end{align}
It is clear that ${{r_{uk}}}$ consists of four parts: the desired signal from the $k$th UE, the compound noise, the inter-UE interference, and the distortion caused by hardware impairments in the UEs' and APs' hardware. It is the last term that makes the analysis in this paper different from prior works, which have assumed perfect hardware. In this section, we will use \eqref{ruk1} to characterize the SE and EE.

\subsection{Spectral Efficiency}

We begin by deriving a closed-form expression for an uplink SE, which is a lower bound on the ergodic capacity. Since the estimate and estimation error are non-Gaussian distributed due to the hardware impairments, we cannot use the standard capacity lower bound from \cite{Hoydis2013a}. The following closed-form SE expression is instead derived using the use-and-then-forget capacity bounding technique \cite{Marzetta2016a}.

\begin{thm} \label{theorem-main}
In cell-free massive  MIMO with hardware impairments, the uplink capacity of the $k$th UE is lower bounded by \vspace{-2mm}
\begin{equation}\label{Rukcf}
R_{uk}  \!=\! \log_2\left( {1 + \frac{{\kappa _r}{\kappa _t}{A}}{ {{\kappa _r}{\text{B}} + {\kappa _r}{\text{C}}- {\kappa _r}{\kappa _t}{A} + {\left( {1 - {\kappa _r}} \right)} {\text{D}} + {{E}}}}} \right),
\end{equation}
where
\begin{align} 
  {\text{A}} &\triangleq {\gamma_k}{\left( {\sum\limits_{m = 1}^M {{\lambda _{mk}}} } \right)^2}, \notag
  \end{align}
\begin{align} 
  {\text{B}} &\triangleq  \sum\limits_{k' =1 }^K  {\gamma_{k'}}  \Bigg( {\sum\limits_{m = 1}^M{\lambda _{mk}}{\beta _{mk'}}} + \rho_p (1-\kappa_r) \sum_{m=1}^M c_{mk}^2 \beta_{mk'}^2 \Bigg), \notag
  \end{align}
  \begin{align} \label{SEpart2}
   {\text{C}} &\triangleq  \sum\limits_{k' =1 }^K  {\gamma_{k'}} \! \left( \left| {{\bm{\varphi }}_k^H{{\bm{\varphi }}_{k'}}} \right|^2 +  \frac{1-\kappa_t}{\kappa_t \tau} \right)\!\! \left( \sum\limits_{m = 1}^M {{\lambda_{mk}}} \frac{\beta_{mk'}}{\beta_{mk}} \right)^2 \!\! \notag,
  \end{align}
  \begin{align}
  {\text{D}} &\triangleq  \sum\limits_{m = 1}^M \Bigg( {{\lambda _{mk}}\sum\limits_{k' = 1}^K {{\gamma_{k'}}{\beta _{mk'}}} }+c_{mk}^2 (1-\kappa_r) \rho_p \beta_{mk'}^2  \notag \\
   &+c_{mk}^2 \kappa_r \rho_p \beta_{mk'} \left(\tau \kappa_t \left| {{\bm{\varphi }}_k^H{{\bm{\varphi }}_{k'}}} \right|^2 +(1-\kappa_t) \right) \Bigg),  \hfill \notag
\end{align}
  \begin{align}
  {\text{E}} &\triangleq \frac{\sigma^2}{\rho _{{u}}} \sum\limits_{m = 1}^M {{\lambda _{mk}}}  \notag\hfill.
\end{align}
\end{thm}
\begin{IEEEproof}
Please refer to Appendix.
\end{IEEEproof}
Theorem~\ref{theorem-main} reveals that the SE increases with the number of APs, which happens when more APs are deployed. The terms $B$ and $C$ in the denominator represents the power of the non-coherent and coherent signals, respectively, from which the desired part $A$ is subtracted. The remainder is interference and the coherent part is due to pilot contamination, caused by pilot reuse and the break of pilot orthogonality by the distortion. The terms $D$ and $E$ represent distortion in the receiving AP and additive noise, respectively. We notice that the SE increases with ${\rho _{u} }$, since it increases the SNR. What is less obvious is that the SE increases with the hardware quality terms $\kappa_t$ and $\kappa_r$, but this will be shown numerically in Section~\ref{se:numerical_result}.

We will now study the SE behavior when we add APs. We assume that the APs are arbitrarily distributed within a finite-sized area, such that ${\beta _{\min }} \leqslant {\beta_{mk}} \leqslant {\beta_{\max }}$ for all $m$, where $0 < {\beta_{\min }}$, ${\beta_{\max }} < \infty $. We then have the following result.

\begin{coro}\label{coro:1}
Suppose the hardware quality factors are replaced by
${\kappa _t} = \frac{{{\kappa _{t0}}}}{{{M^{{z_t}}}}}$, ${\kappa _r} = \frac{{{\kappa _{r0}}}}{{{M^{{z_r}}}}}$, for some constant ${\kappa _{{\text{t0}}}}, {\kappa _{{\text{r0}}}}>0$, where $z_{t}$, $z_{r}$ denote scaling exponents in transmitters and receivers, respectively. If $z_t>0$ and $z_r \geq 0$ (or $z_t=0$ and $z_r > 1/2$), then
\begin{equation} \label{law_1}
R_{uk}  \to 0, \quad \text{as } M \to \infty.
\end{equation}
If instead $z_t = 0$ and $0 < z_r < 1/2$, then $R_{uk}  \to {\log _2}\left( 1 + \mathrm{SIR}_{k}^{\infty} \right)$ as $M \to \infty$, where
\begin{equation} \label{law_2}
\mathrm{SIR}_{k}^{\infty} =
\frac{ \kappa_{t0}  }{  \!
\sum\limits_{k' =1 }^K \! \frac{\gamma_{k'}}{\gamma_k} \! \left( \left| {{\bm{\varphi }}_k^H{{\bm{\varphi }}_{k'}}} \right|^2 +  \frac{1-\kappa_{t0}}{\kappa_{t0} \tau} \right)\!\! \frac{\left( \sum\limits_{m = 1}^M {{\mu_{mk}}} \frac{\beta_{mk'}}{\beta_{mk}} \right)^{\!2}}{\left( \sum\limits_{m}^{M} \mu_{mk} \right)^{\!2}} - \kappa_{t0} }
\end{equation}
and $\mu_{mk} = \frac{\rho_p \beta_{mk}^2}{\rho_p \beta_{mk} + \sigma^2}$.
\end{coro}
\begin{IEEEproof}
We  divide all  terms in \eqref{Rukcf} by ${\kappa _r}{\kappa _t}{A}$ to obtain
\begin{equation}
R_{uk}  \!=\! \log_2\left( {1 + \frac{ 1}{ { \frac{ {{B}}}{{\kappa _t}{A}} + \frac{ {{C}}}{{\kappa _t}{A}}  - 1 + \frac{ {\left( {1 - {\kappa _r}} \right)} {{D}} }{{\kappa _r}{\kappa _t}{A}} + \frac{{E}}{{\kappa _r}{\kappa _t}{A}}  }}} \right)\notag.
\end{equation}
Under the assumption that all $\beta_{mk}$ are strictly non-zero and bounded, it is straightforward to show that  $\frac{ {{C}}}{{\kappa _t}{A}}  - 1 = {\rm\emph{O}}\left(M^{2z_t}\right)$, which implies that $R_{uk} \to 0$ unless $z_t=0$.  We further notice that $ \frac{ {{B}}}{{\kappa _t}{A}} +  \frac{ {\left( {1 - {\kappa _r}} \right)} {{D}} }{{\kappa _r}{\kappa _t}{A}} + \frac{{E}}{{\kappa _r}{\kappa _t}{A}} = {\rm\emph{O}}\left(M^{2z_r-1}\right)$ when $z_t=0$ and $z_r \geq0$, thus these terms vanish asymptotically if
$2z_r-1 < 0$ or $z_r < 1/2$. In contrast, if $z_r > 1/2$, these terms grow unboundedly and $R_{uk} \to 0$. This proves \eqref{law_1}.

In the case of $z_t=0$ and $0 < z_r < 1/2$, all term in the denominator vanishes, except ``$-1$'' and
\begin{equation}
\frac{ {{C}}}{{\kappa _t}{A}}  \to \frac{
\sum\limits_{k' =1 }^K  {\gamma_{k'}} \! \left( \left| {{\bm{\varphi }}_k^H{{\bm{\varphi }}_{k'}}} \right|^2 +  \frac{1-\kappa_{0t}}{\kappa_{0t} \tau} \right)\!\! \left( \sum\limits_{m = 1}^M {{\mu_{mk}}} \frac{\beta_{mk'}}{\beta_{mk}} \right)^2 }{ \kappa_{0t} \gamma_k \left( \sum\limits_{m}^{M} \mu_{mk} \right)^2}.\notag
\end{equation}
This leads to the asymptotic expression in \eqref{law_2}.
\end{IEEEproof}

Corollary~\ref{coro:1} proves that the APs can tolerate much lower hardware quality as the number of APs increases. However, the hardware quality of the UEs cannot be reduced without suffering a substantial performance loss. This is an important result for practical deployment of cell-free massive MIMO systems, since it indicates that low-cost AP hardware can be used.  Note that a similar result has been shown for cellular massive MIMO in \cite{bjornson2015massive,Bjornson2014Massive}, but for co-located arrays with many antennas, which is a very different topology.

\subsection{Energy Efficiency}
In the following, we consider the EE of cell-free massive MIMO systems to see how it is affected by the number of APs. The EE (bit/Joule) is defined as the ratio of the sum rate (bit/s) to the total power consumption (Watt) of the system. As in \cite{ngo2017total}, we consider a realistic power consumption model where the total power consumption includes the power consumption of the transmitters, receivers, and backhaul. More precisely, the total power consumption is modeled as
 \begin{eqnarray}\label{power}
{P_\textrm{total}} = \sum\limits_{k = 1}^K {{P_k}}  + \sum\limits_{m = 1}^M {{P_m}}  + \sum\limits_{m = 1}^M {{P_{b,m}}} ,
 \end{eqnarray}
where ${{P_m}}$ denotes the circuit power consumption at the $m$th AP (including analog transceiver components and digital signal processing), ${{P_{b,m}}}$ is the power consumed by the backhaul link connecting CPU and the $m$th AP, and ${P_k}$ is the power consumption at the $k$th UE (including the radiated transmission power, amplifier inefficiency and the circuit power). Then, the EE can be expressed as
\begin{eqnarray}\label{EE}
{\text{EE = }}{\raise0.7ex\hbox{${\sum\limits_{k = 1}^K {{R_{uk}} \cdot B} }$} \!\mathord{\left/
 {\vphantom {{\sum\limits_{k = 1}^K {{R_{uk}} \cdot B} } {\left( {\sum\limits_{k = 1}^K {{P_k}}  + \sum\limits_{m = 1}^M {{P_m}}  + \sum\limits_{m = 1}^M {{P_{b,m}}} } \right)}}}\right.\kern-\nulldelimiterspace}
\!\lower0.7ex\hbox{${\left( {\sum\limits_{k = 1}^K {{P_k}}  + \sum\limits_{m = 1}^M {{P_m}}  + \sum\limits_{m = 1}^M {{P_{b,m}}} } \right)}$}},
\end{eqnarray}
where $B$ denotes the bandwidth.

\section{Numerical Results}\label{se:numerical_result}
In this section, we study the SE and EE numerically. We assume that the $M$ APs and $K$ UEs are independently and uniformly distributed within a square of size $1 \times 1$ k$\text{m}^2$. The variance $\beta_{mk}$ in (\ref{gmk}) is computed as
\begin{eqnarray}\label{Bmk}
{\beta _{mk}} = {L_{mk}^{ - \alpha }} \cdot {10^{\frac{{z_{mk}}}{{10}}}},
\end{eqnarray}
where $L_{mk}$ (km) is the distance between the $k$th UE and $m$th AP, $\alpha$ is the path loss exponent, and ${z_{mk}} \sim \mathcal{N}\left( {0,{\sigma_{sh}^2}} \right)$ is the shadow fading.
We also use the simulation parameters summarized in Table \ref{parameter}. 
The noise variance is computed as $\sigma^2 = B \cdot {k_B} \cdot {T_0} \cdot {\text{noise figure }}\left( {\text{W}} \right)$, where ${k_B} = 1.381 \cdot {10^{ - 23}}\left( {{\text{Joule per Kelvin}}} \right)$, and  ${T_0} = 290\left( {{\text{Kelvin}}} \right)$.

\begin{table}[tb]
\renewcommand{\thetable}{\Roman{table}}
\caption{Key Simulation Parameters}
\label{parameter}
\centering
\begin{tabular}{|c|c|}
\hline
Parameters  &  Values\\
\hline
noise figure&  9 dB   \\
\hline
$B$ &  20 MHz   \\
\hline
$\rho_p$, $\rho_u$ &  100 mW   \\
\hline
${\sigma _{sh}}$ &  8 dB  \\
\hline
$\alpha $ &  3.5   \\
\hline
${\gamma _k}$ &  1  \\
\hline
\end{tabular} \vspace{-3mm}
\end{table}

\begin{figure}[t]
\centering
\includegraphics[scale=0.42]{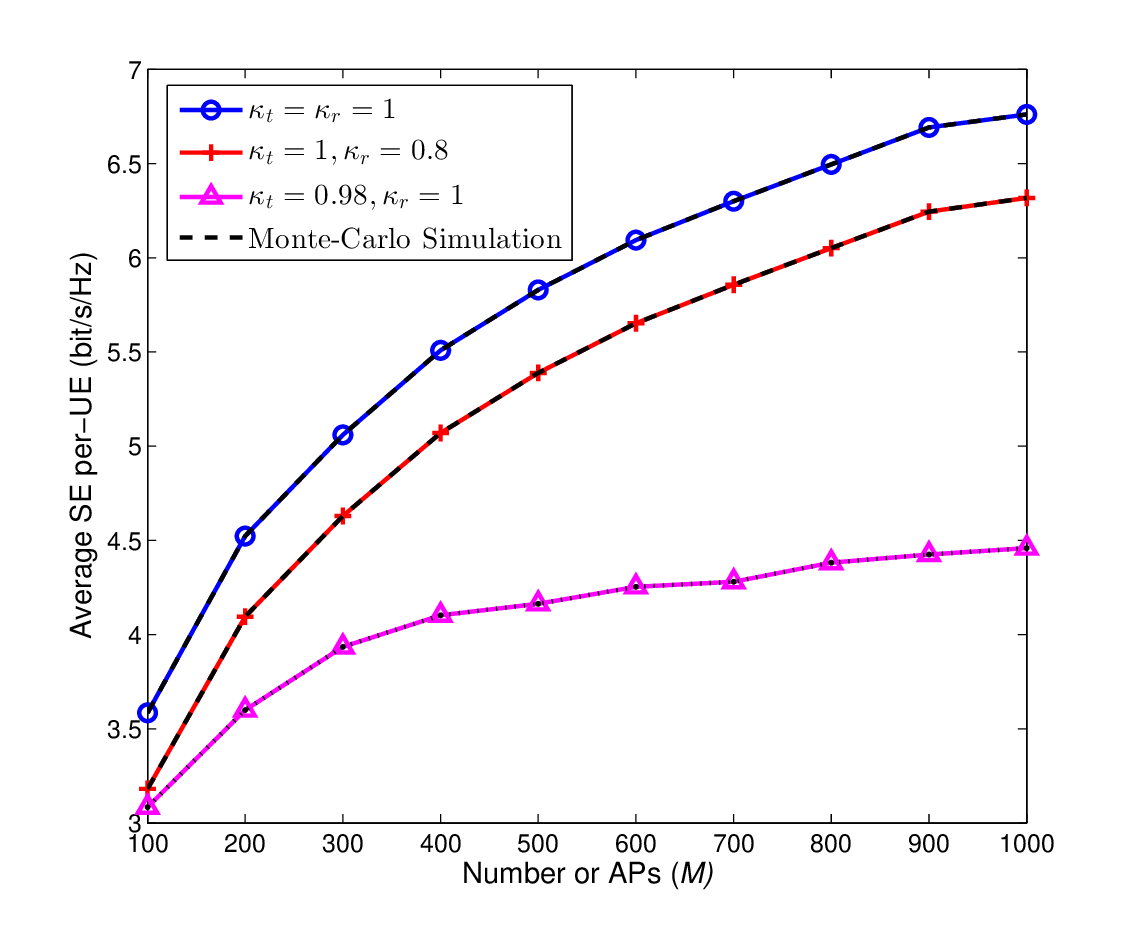} \vspace{-3mm}
\caption{Average SE per-UE as a function of the number of APs for $K = 10$. Here, ${\tau} = K$ and all pilot sequences are pairwise orthogonal.}
\label{aveSE} \vspace{-4mm}
\end{figure}
The Monte Carlo simulated and analytical average SE in  (\ref{Rukcf}) are compared in  Fig.~\ref{aveSE}, as a function of the number of APs. It is clear to see that the analytical and simulated curves are almost the same for all considered cases. The average SE is an increasing function of $M$. The SE decreases when the hardware qualities $\kappa_t$ and $\kappa_r$ decrease. Nevertheless, Fig.~\ref{aveSE} shows that the SE is mainly limited by the hardware impairments at the UE (e.g., $\kappa_t=0.98$ gives a larger impact than $\kappa_r=0.98$). 
\begin{figure}[t]
\centering
\includegraphics[scale=0.41]{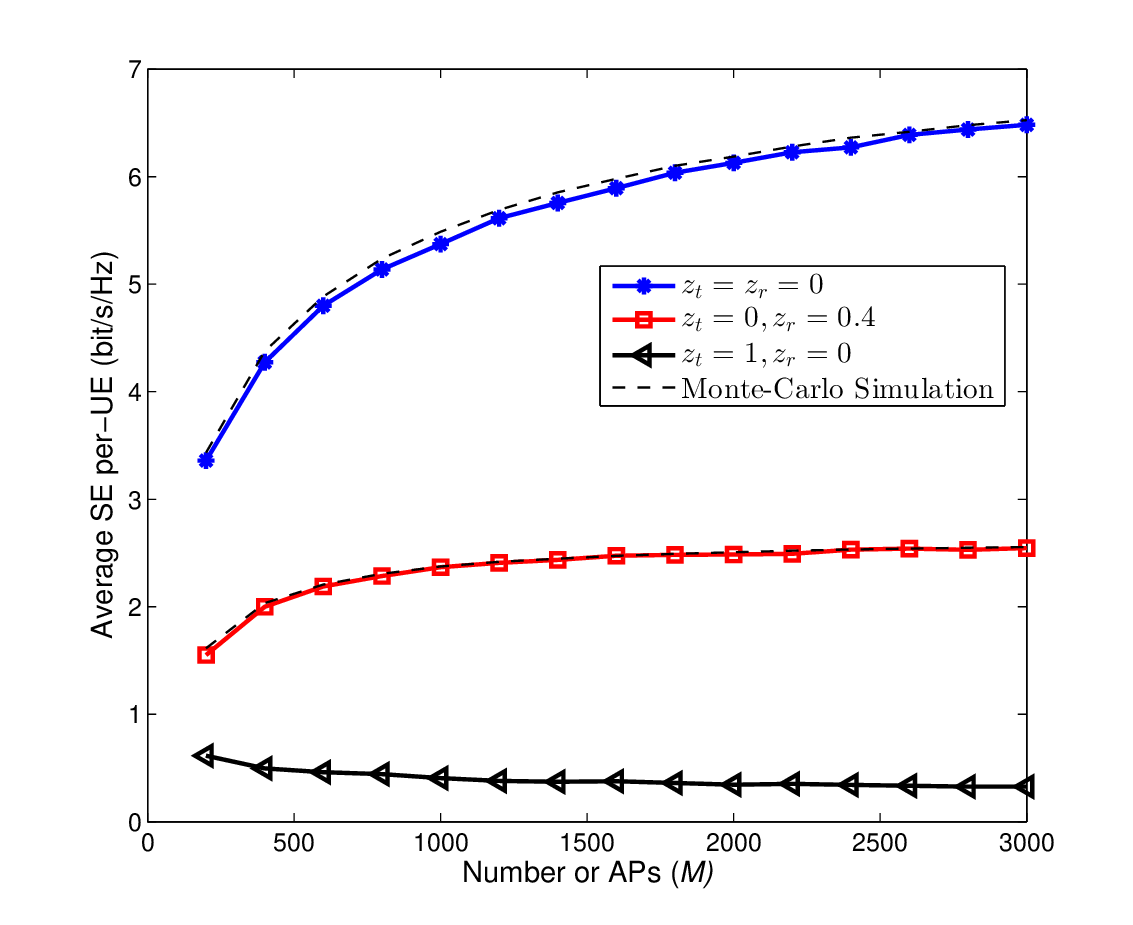} \vspace{-4mm}
\caption{Average SE per-UE against the number of APs for different hardware scaling factors $z_t$, $z_r$.}
\label{SumSE} \vspace{-4mm}
\end{figure}

Fig.~\ref{SumSE} validates the hardware-quality scaling law established by Corollary 1. When $z_t=z_r=0$, the SE increases with $M$ without bound. When $z_t=0,0\leq z_r<1/2$ (e.g., $z_r=0.4$), we observe that the SE converges to a non-zero limit. Moreover, the SE converges to zero when $z_r=0,z_t>0$.

\begin{figure}[t]
\centering
\includegraphics[scale=0.42]{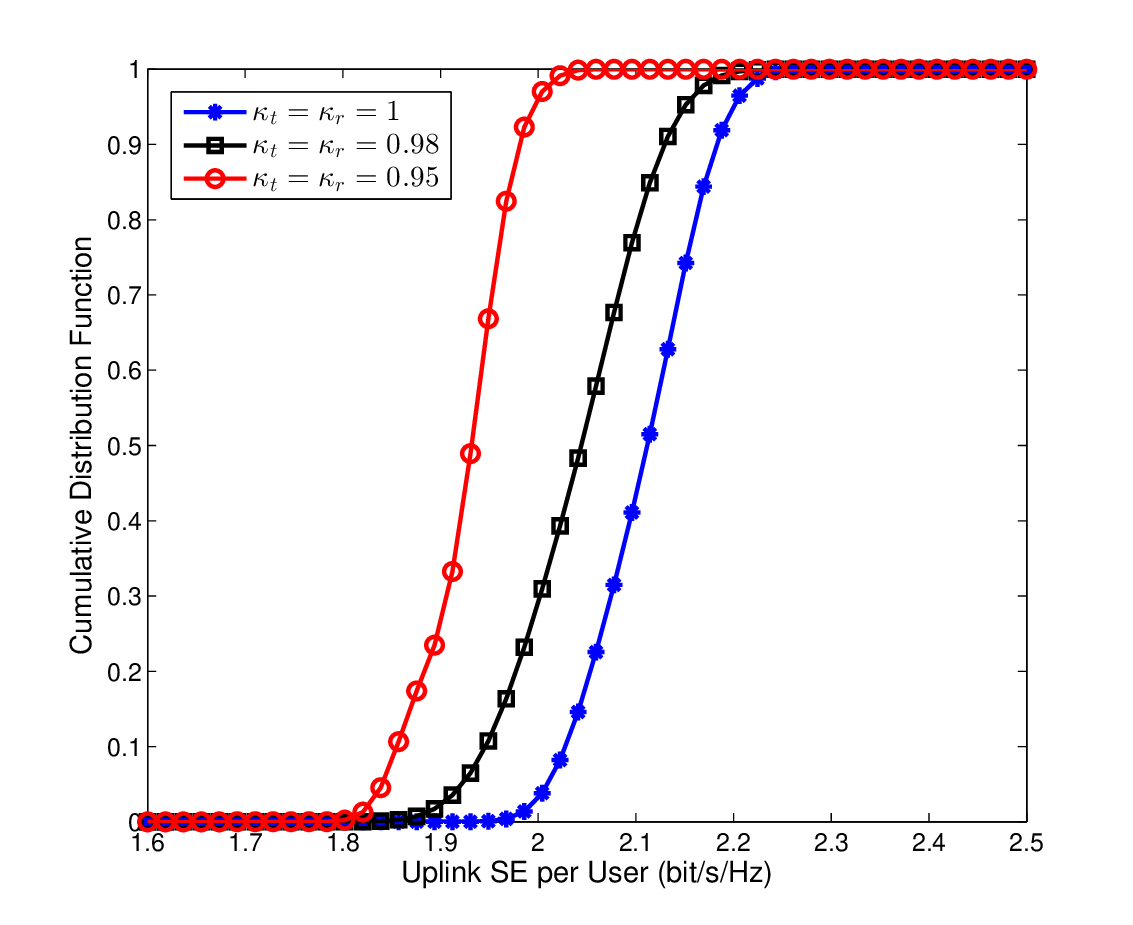} \vspace{-3mm}
\caption{SE CDF for different levels of hardware impairments ${\kappa_{{t}}}$ and ${\kappa_{ {r}}}$ $\left(M = 200, K = 60\right)$.}
\label{CDF} \vspace{-4mm}
\end{figure}

Fig.~\ref{CDF} presents the CDF of the per-UE SE with $M = 200$, $K = 60$, and ${\tau  } = 20$, for different hardware qualities $\kappa_t$ and $\kappa_r$. We find that around $80\%$ of the SE values are distributed between $1.96$ and $2.2$ for ${\kappa_{t} =\kappa_{r} =1}$, while the range is $1.89-2.1$ for ${\kappa_{t} =\kappa_{r}=0.98}$ and $1.8-1.92$ for ${\kappa_{t} =\kappa_{r} =0.95}$. Hence, the uplink per-UE SE for ${\kappa_{t} =\kappa_{r} =1}$ is only $5\%$ higher than in the case when  ${\kappa_{t} =\kappa_{r} =0.95}$.

\begin{figure}[t]
\centering
\includegraphics[scale=0.408]{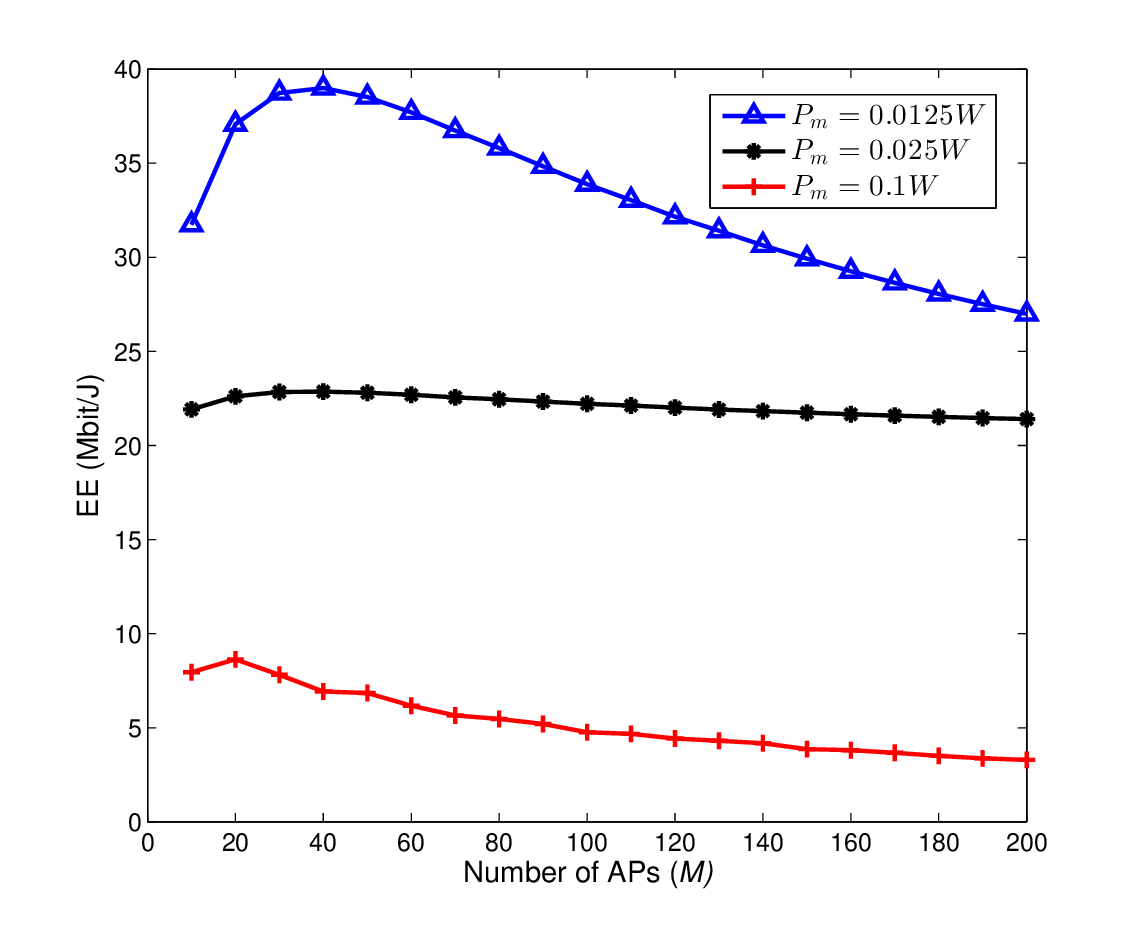} \vspace{-3mm}
\caption{EE as a function of the number of APs for different power consumption $P_m$ ($K = 20$, ${P_k} = 0.6$\,W, and ${P_{b,m}} = 0.1$\,W).}
\label{EEMchange} \vspace{-4mm}
\end{figure} 

Fig.~\ref{EEMchange} investigates the EE in \eqref{EE} as a function of the number of APs for different values of $P_m$. We consider $K = 20$, ${P_k} = 0.6$\,W,  and ${P_{b,m}} = 0.1$\,W. We observe that the EE decreases when increasing $P_m$, for the same number of APs, due to the larger power consumption. There is a value ${M^{\text{opt}}}$ that provides maximum EE. For $P_m=0.0125$, this value is ${M^{\text{opt}}}=40$. When $M \leqslant {M^{\text{opt}}}$, the EE can be improved by increasing $M$. However, when $M > {M^{\text{opt}}}$, increasing $M$ will rapidly reduce the EE. This is due to the fact that only a few APs have a large impact on the SE of a UE, thus adding more APs will increase the power consumption linearly, while the sum SE might increase more slowly.

\vspace{-2mm}
\section{Conclusion}\label{se:conclusion}
This paper has taken a first look at the impact of transceiver hardware impairments on the performance of cell-free massive MIMO systems, using a well-established distortion model. Closed-form expressions for the SE and EE were obtained, which reveal how the performance depends on the hardware quality factors of the APs and UEs, the number of APs $M$, and the number of UEs $K$. Furthermore, a hardware-quality scaling law was established. It proves that the detrimental effect of hardware impairments at the APs vanishes when the number of APs  grows large (in a finite-sized deployment area), while the effect of hardware impairments at the UEs remain. This indicates that cell-free massive MIMO can be deployed using low-quality hardware. In future work, more detailed and specialized hardware impairment models can be used to validate these observations.

\vspace{-2mm}
\section{Appendix}\label{se:appendix}
The received signal ${r_{uk}}$ in (\ref{ruk1}) can be rewritten as
\begin{align}\label{ruk2}
{r_{{{u}}k}}& = {\text{D}}{{\text{S}}_k} \cdot {q_k} + {\text{B}}{{\text{U}}_{k}} \cdot {q_k} + \sum\limits_{k' \ne k}^K {{\text{U}}{{\text{I}}_{kk'}} \cdot {q_{k'}}}  \hfill \notag\\
&+ \sum\limits_{k' = 1}^K {{\text{HI}}_{{{tkk'}}}^{{\text{UE}}}}  + {\text{HI}}_{{r}}^{{\text{AP}}} + {\text{N}}{{\text{I}}_k},
\end{align}
where
\begin{align}
{\text{D}}{{\text{S}}_k} &\triangleq \sqrt {\rho _{{u}} {\gamma_k}{{\kappa _r}{\kappa _t}}} {\rm \mathbb{E}}\left\{ {\sum\limits_{m = 1}^M {\hat g_{mk}^*{g_{mk}}} } \right\},\hfill \notag
\end{align}
\begin{align}
  {\text{B}}{{\text{U}}_{k}} &\triangleq \sqrt {\rho _{{u}} {\gamma_k}{{\kappa _r}{\kappa _t}}} \left( {\sum\limits_{m = 1}^M {\hat g_{mk}^*{g_{mk}}}  \!-\! {\rm\mathbb {E}}\left\{ {\sum\limits_{m = 1}^M {\hat g_{mk}^*{g_{mk}}} } \right\}} \right), \hfill \notag
\end{align}
\begin{align}
  {\text{U}}{{\text{I}}_{kk'}} &\triangleq \sqrt {\rho _{{u}} {{\kappa _r}{\kappa _t}} {\gamma_{k'}} } \sum\limits_{m = 1}^M { \hat g_{mk}^*{g_{mk'}}},   \hfill  \notag
\end{align}
\begin{align}
  {\text{HI}}_{{tkk'}}^{{\text{UE}}} &\triangleq \sum\limits_{m = 1}^M {{\sqrt {{\kappa _r}} }\hat g_{mk}^*{g_{mk'}}{\eta _{k't}}}, \hfill \notag
\end{align}
\begin{align}
{\text{HI}}_{{r}}^{{\text{AP}}} \triangleq \sum\limits_{m = 1}^M {\hat g_{mk}^*{\eta _{mr}}},\hfill \notag
  {\text{N}}{{\text{I}}_k} \triangleq \sum\limits_{m = 1}^M {\hat g_{mk}^*{w_{um}}}.  \notag\hfill
\end{align}

By using the use-and-then-forget bounding technique \cite[Chapter 3]{Marzetta2016a}, the achievable SE of the $k$th UE is obtained as
\begin{align}\label{Rcf}
{\text{R}}_{uk}  = \log_2\left( {1 + \frac{{{{\left| {{\text{D}}{{\text{S}}_k}} \right|}^2}}}{{{\rm \mathbb{E}}\left\{ {{{\left| {{\text{B}}{{\text{U}}_{kk'}}} \right|}^2}} \right\} + \sum\limits_{k' \ne k}^K {{\rm \mathbb{E}}\left\{ {{{\left| {{\text{U}}{{\text{I}}_{kk'}}} \right|}^2}} \right\} + \sum\limits_{k' = 1}^K {{\rm \mathbb{E}}\left\{ {{{\left| {{\text{HI}}_{ {tkk'}}^{{\text{UE}}}} \right|}^2}} \right\} + {\rm \mathbb{E}}\left\{ {{{\left| {{\text{HI}}_{ {r}}^{{\text{AP}}}} \right|}^2}} \right\} + {\rm \mathbb{E}}\left\{ {{{\left| {{\text{N}}{{\text{I}}_k}} \right|}^2}} \right\}} } }}} \right).
\end{align}

It is straightforward, but tedious, to compute the following expectations:

\begin{equation}
 {{{\left| {\text{D}}{{\text{S}}_k}\right|}^2}}  = \rho _{{u}} {\gamma_k}{{\kappa _r}{\kappa _t}} \left( \sum\limits_{m = 1}^M {{\lambda_{mk}}} \right)^2  \hfill \label{EDS},
 \end{equation}
 \begin{align} \notag
  \mathbb{E}\left\{ {{{\left| {{\text{B}}{{\text{U}}_k}} \right|}^2}} \right\} &= \rho _u{\gamma_k}{{\kappa _r}{\kappa _t}} \Bigg( \sum\limits_{m = 1}^M {{\gamma _{mk}}{\beta _{mk}}}  + \frac{1-\kappa_t}{\kappa_t \tau} \! \! \left( \sum\limits_{m = 1}^M {{\lambda_{mk}}} \right)^2  \\ & + \rho_p (1-\kappa_r) \sum_{m=1}^M c_{mk}^2 \beta_{mk}^2  \Bigg) \hfill \label{EBU},
  \end{align}
  \begin{align}
  & \sum\limits_{k' \ne k}^K {\mathbb{E}\left\{ {{{\left| {{\text{U}}{{\text{I}}_{kk'}}} \right|}^2}} \right\}}
  =\rho _{{u}} {{\kappa _r}{\kappa _t}}\sum\limits_{k' \ne k}^K  {\gamma_{k'}} \Omega_{kk'}
  \label{EUI} ,
\end{align}
    where
\begin{align} \notag
      \Omega_{kk'} \triangleq  & \mathbb{E} \left\{ \left| \sum\limits_{m = 1}^M { \hat g_{mk}^*{g_{mk'}}}  \right|^2 \right\} =  \Bigg( {\sum\limits_{m = 1}^M{\lambda _{mk}}{\beta _{mk'}}}  \\
  & + \left( \left| {{\bm{\varphi }}_k^H{{\bm{\varphi }}_{k'}}} \right|^2 +  \frac{1-\kappa_t}{\kappa_t \tau} \right) \left( \sum\limits_{m = 1}^M {{\lambda_{mk}}} \frac{\beta_{mk'}}{\beta_{mk}} \right)^2 \notag \\ & + \rho_p (1-\kappa_r) \sum_{m=1}^M c_{mk}^2 \beta_{mk'}^2 
  \Bigg).
    \end{align}
 This expression is also utilized to compute
\begin{align}
  &\sum\limits_{k' = 1}^K {{\mathbb{E}}\left\{ {{{\left| {{\text{HI}}_{ {t}kk'}^{{\text{UE}}}} \right|}^2}} \right\}}  =\rho _{{u}} {\kappa _r}(1-{\kappa _t})
  \sum\limits_{k' = 1}^K {\gamma _{k'}} \Omega_{kk'}  \hfill,\label{EHIt}
\end{align}

  \begin{align} \notag
  &{\mathbb{E}}\left\{ {{{\left| {{\text{HI}}_{ {r}}^{{\text{AP}}}} \right|}^2}} \right\}
   =(1-{\kappa _r})\rho _{{u}} \sum\limits_{m = 1}^M \Bigg( {{\lambda _{mk}}\sum\limits_{k' = 1}^K {{\gamma_{k'}}{\beta _{mk'}}} }  \\ \notag
   &+c_{mk}^2 \kappa_r \rho_p \beta_{mk'} \left(\tau \kappa_t \left| {{\bm{\varphi }}_k^H{{\bm{\varphi }}_{k'}}} \right|^2 +(1-\kappa_t) \right) \\
   &+c_{mk}^2 (1-\kappa_r) \rho_p \beta_{mk'}^2 
   \Bigg) ,\label{EHIr}
\end{align}

  \begin{equation}
  \mathbb{E}\left\{ {{{\left| {{\text{N}}{{\text{I}}_k}} \right|}^2}} \right\} = \sigma^2 \sum\limits_{m = 1}^M {{\lambda_{mk}}}  \hfill. \label{ENI}
\end{equation}

The proof is completed by substituting (\ref{EDS})--(\ref{ENI}) into (\ref{Rcf}).
\bibliographystyle{IEEEtran}
\bibliography{IEEEabrv,Ref}
\end{document}